# Nuclear ashes and outflow in the eruptive star *Nova Vul 1670*

Tomasz Kamiński[1,2], Karl M. Menten[2], Romuald Tylenda[3], Marcin Hajduk[3], Nimesh A. Patel[4], Alexander Kraus[2]

[1]European Southern Observatory, Alonso de Córdova 3107, Vitacura, Santiago, Chile
[2]Max-Planck Institut für Radioastronomie, Auf dem Hügel 69, 53121 Bonn, Germany
[3]Department for Astrophysics, N. Copernicus Astronomical Center, Rabiańska 8, 87-100, Toruń, Poland
[4]Harvard-Smithsonian Center for Astrophysics, 60 Garden Street, Cambridge, MA 02138, USA

**CK Vulpeculae was observed in outburst in 1670-1672[1], but no counterpart was seen until 1982, when a bipolar nebula was found at its location[1–3]. Historically, CK Vul has been considered to be a nova (*Nova Vul 1670*), but a similarity to 'red transients', which are more luminous than classical nova and thought to be the result of stellar collisions[4], has re-opened the question of CK Vul's status[5–6]. Red transients cool to resemble late M-type stars, surrounded by circumstellar material rich in molecules and dust[7–9]. No stellar source has been seen in CK Vul, though a radio continuum source was identified at the expansion centre of the nebula[3]. Here we report CK Vul is surrounded by chemically rich molecular gas with peculiar isotopic ratios, as well as dust. The chemical composition cannot be reconciled with a nova or indeed any other known explosion. In addition, the mass of the surrounding gas is too high for a nova, though the conversion from observations of CO to a total mass is uncertain. We conclude that CK Vul is best explained as the remnant of a merger of two stars.**

Using the submillimetre-wave APEX telescope we discovered bright and chemically complex molecular gas in emission which has never been observed in CK Vul before. A spectral-line survey in the 217–910 GHz range revealed emission from a plethora of molecules (Fig. 1a–c). Additionally, using the Effelsberg radio telescope, we observed inversion lines of $NH_3$ (Fig. 1d). The detected transitions are listed in Extended Data Table 1. Our excitation analysis indicates that the molecular gas is cool, with rotational temperatures of 8–22 K, but some amount of gas at higher excitation is also evident.

The molecular inventory implies that the abundance of nitrogen is greatly enhanced in CK Vul. The paucity of oxides, namely the lack of SO, $SO_2$, and maser emission of $H_2O$, and OH – typically omnipresent in *oxygen-rich* environments – implies that the circumstellar material is not dominated by oxygen. We do observe relatively strong lines from some species containing oxygen, i.e. SiO, CO, $HCO^+$, and $H_2CO$, but those molecules are also observed in envelopes of carbon-rich stars. The gas does not appear to be *carbon-rich* because many species typical for such environments (e.g. SiC, $SiC_2$, and $HC_3N$), although covered by our spectra, are not observed. Again, all the carbon-bearing molecules present in CK Vul, i.e. CO, CS, $H_2CO$, and $HCO^+$, have been observed in other chemical types of circumstellar envelopes[10]. Unusual in CK Vul is a rich variety of nitrogen-bearing species. From all nitrogen-bearing species predicted in thermal-equilibrium to be abundant in gas greatly enhanced in nitrogen[11], only NO and $N_2$ remain undetected in CK Vul. The $N_2$ molecule has no allowed rotational transitions, while transitions of NO, although covered by APEX spectra, might have been undetected owing to the low levels of oxygen and a small dipole moment of NO molecule. All the nitrogen-bearing species observed in CK Vul are also present in the envelopes of the yellow supergiant IRC+10420 and the luminous blue variable η Carinae which both were recently proposed to be prototypes of *nitrogen-rich* objects[11,12]. This makes CK Vul only the third known such case. An overabundance of nitrogen relative to oxygen was suggested before for CK Vul based on observations of the optical atomic lines[1], but the result was questionable owing to uncertain assumptions.

Some of the transitions covered by APEX were later observed at higher angular resolution with the Submillimeter Array (SMA). The maps reveal that the emission arises from a bipolar structure ~15" in size. This molecular region is much smaller than the long-known ionised nebula (extent of ~71"; Fig. 2a). The spatio-kinematical structure of the lobes is complex and suggests the presence of two partially overlapping hour-glass shells observed at very low inclination angles. The lobes are apparent only in some of the observed transitions; most of the molecular emission arises in the central source which is only partially resolved at our best resolution of ~2". The northern molecular lobe coincides very closely with the brightest clump of the optical nebula (Fig. 2b), suggesting that the molecular gas coexists with the plasma. The observed misalignment of the long axis of the molecular region with respect to the axis of the large-scale optical nebula might be caused by precession.

In addition to molecular lines, continuum emission was observed with the SMA revealing thermal emission of dust arising from the same position where radio continuum was found in earlier observations[3]. The millimetre-wave source is dominated by a structure 3.7"×1.0" in size and seen at the position angle of 33°.4, but has also components extending a few arcseconds along the northern and southern molecular lobe. The continuum indicates the presence of a flattened dusty envelope, perhaps a torus, and a pair of collimated jets. Our analysis of all available continuum measurements ranging from micrometre to centimetre wavelengths indicates that the emission is dominated by dust at a temperature of ~15 K but warmer dust up to 50 K must also be present.

From our rough estimate of the CO column density, $N(CO)=4 \cdot 10^{17}/cm^2$, we calculate the total mass of the gas to be ~1 $M_\odot$. Here we assumed that the CO abundance with respect to hydrogen is of the order of $10^{-4}$ as found in many interstellar/circumstellar environments of various types. The possible line-saturation effects would make our estimate a lower limit on the total mass. The peculiar elemental composition of CK Vul indicates, however, that the CO abundance may deviate from the classical value. If the overabundance of nitrogen is owing to its production at the cost of carbon and oxygen, the actual CO abundance with respect to hydrogen can be lower than $10^{-4}$ and then our value underestimates the total mass of the gas. In case of strong enrichment of helium on the cost of hydrogen, our estimate should be reasonably close to the actual total mass, as the correction for the presence of helium would compensate the deficiency of hydrogen. Also, the mass should be enlarged by the contribution of the material seen in the optical nebula, a number which remains unknown. Our mass estimate, although uncertain, is much higher than what classical-nova explosions can accumulate during their life-time[13].

The presence of the strong submillimetre-wave molecular emission itself makes CK Vul an extraordinary eruptive variable. Classical novae do not show such emission as we recently confirmed by observing 17 Galactic-disk novae with APEX. Galactic red transients, which have rich molecular spectra at optical and near-infrared wavelengths, have neither been detected in submillimetre-wave thermally-excited emission lines[8,14].

In fact, the central object of CK Vul may be hostile to molecules, as suggested by the presence of the ionic species $HCO^+$ and $N_2H^+$. Their formation channels in the absence of water require a high abundance of $H_3^+$ which can be formed from $H_2$ exposed to an ultraviolet radiation field[15] or by shocks. The high outflow velocity of ~210 km/s observed in CK Vul, the presence of jets, and emission of atomic ions[3,16] make shocks a more favourable ionisation mechanism.

There is a striking resemblance of the newly-revealed observational characteristics of CK Vul to a short stage of low- to intermediate-mass stars known as preplanetary nebulae (PPNe), especially to OH231.8+4.2 (a.k.a. Calabash Nebula) which has an extended pair of lobes seen in optical atomic lines and a pair of molecular jets emanating from a dusty flattened structure[17]. At least some of the known PPNe must have been formed in a short and energetic event[18,19] and it has recently been proposed that the type of explosions we have witnessed in red transients may be actually responsible for the formation of the circumstellar material of PPNe[18,20]. Our observations of CK Vul would then provide strong support for such a link.

However, our analysis leads to the conclusion that the remnant of *Nova 1670* must be of a different nature than PPNe. First of all, its spectral energy distribution (Extended Data Fig. 1) implies a luminosity of ≳0.9 $L_\odot$, while PPNe reach luminosities of the order of $10^4\,L_\odot$. Moreover, the chemical composition of CK Vul, especially the nitrogen enrichment, would be very unusual for a PPN. Anomalous is also the presence of lithium in the outflow of the oldest *nova* as evidenced by two variable field stars whose spectra show absorption lines of lithium[16].

The strongest argument for CK Vul being indeed a truly unique transient comes from our analysis of its isotopic abundances. The column density ratios of different isotopologues listed in Table 1 likely represent the true isotopic ratios of the different elements (but may be somewhat influenced by photo-chemical fractionation and opacity effects). The isotopic ratios of the CNO elements compared to solar values[21] (in brackets), i.e. $^{12}C/^{13}C$=2-6 (89), $^{14}N/^{15}N$≈26 (272), $^{16}O/^{18}O$≈23 (499), and $^{16}O/^{17}O$>225 (2,682), reveal a very peculiar isotopic pattern which undoubtedly indicates nuclear processing of the circumstellar gas. The pattern could not be produced by an asymptotic-giant-branch (AGB) star or post-AGB/PPN object because these are characterised by much higher ratios of $^{16}O/^{18}O$ and $^{14}N/^{15}N$ (ref. 22); in fact, isotopologues containing $^{15}N$ are never observed in spectra of those evolved stars. The obtained isotopic ratios cannot be reconciled with the current understanding of thermonuclear runaway nucleosynthesis, mainly because nova ashes have a much lower $^{16}O/^{17}O$ ratio[23].

It is most tempting to consider that CK Vul underwent its XVII-century cataclysm due to a merger of stars, as such events are now proven to explain the explosions of red transients[4]. The explosion could have been violent enough to penetrate and eject inner parts of the merging stars, exposing material that was active in nuclear burning. Interestingly, the general elemental abundances revealed by the molecular spectra here are well reproduced by abundances expected for non-explosive hydrogen burning in the CNO cycles[24]. Not all of the observed isotopic signatures fit those models, though, with the observed ratios of $^{15}N/^{14}N$ and $^{16}O/^{18}O$ being too high. However, a merger remnant could be a complex mixture of processed and unprocessed gas and no quantitative predictions exists for the chemical composition of such an exotic star and its circumstellar environment.

Interestingly, the $^{12,13}C$ and $^{14,15}N$ isotopic ratios of CK Vul are close to those of presolar grains known as *nova grains*[25] but of unclear origin[26,27]. Although the agreement in abundances is conspicuous, those stardust grains originate from carbon-rich environments, what makes their link to the CK-Vul phenomenon elusive.

**Acknowledgements** We thank F. Wyrowski, A. Belloche, T. Csengeri, K. Immer, K. Young and the APEX staff for executing part of the observations reported here. The Atacama Pathfinder Experiment (APEX) is a collaboration between the Max-Planck-Institut für Radioastronomie, the European Southern Observatory, and Onsala Space Observatory. The SMA is a joint project between the Smithsonian Astrophysical Observatory and the Academia Sinica Institute of Astronomy and Astrophysics. We thank the SMA director R. Blundell for granting us director's discretionary time. The Effelsberg 100-m radio telescope is operated by the Max-Planck-Institut für Radioastronomie on behalf of the Max-Planck-Gesellschaft.

**Figure 1 | Spectra of CK Vul with identifications of the main features. a-c**, Example APEX spectra containing some of the observed emission lines. Overall lines of CO, CS, SiO, CN, HCN, HNC, HCO$^+$, N$_2$H$^+$, H$_2$CO, and their isotopologues were observed. **d**, The Effelsberg spectrum of the inversion lines of ammonia. The rotational lines show very broad profiles with full widths of up to ~420 km/s.

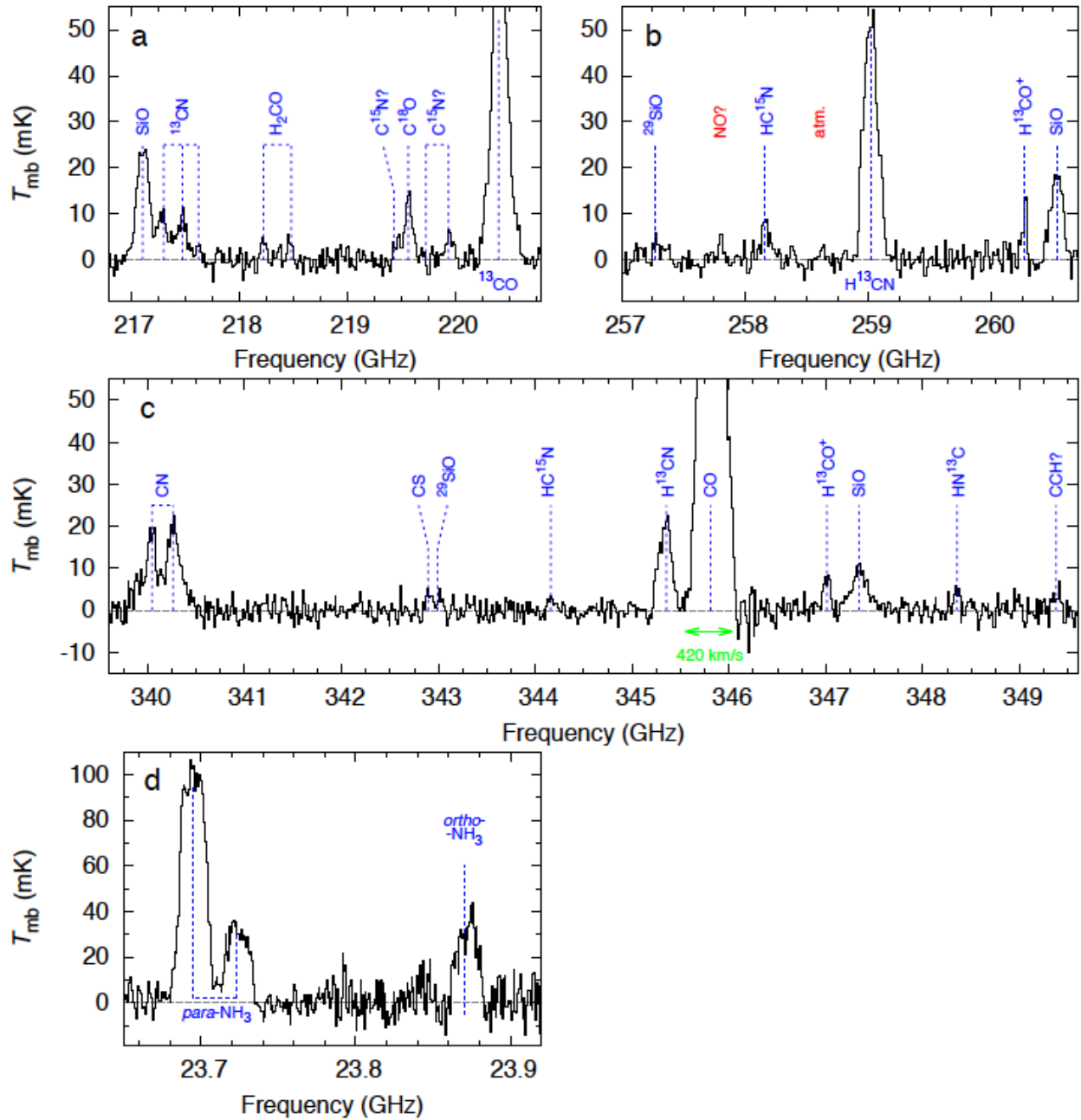

**Figure 2 | The ionised nebula and the newly discovered molecular emission in CK Vul**. **a**, The image shows the Hα+[NII] nebula created in the XVII-century explosion. Bright stars were removed from this optical image[3]. Green contours show the emission in the $^{12}$CO $J$=3–2 transition observed at submillimetre wavelengths (at 29, 43, 57, 72, 86% of the maximum emission). **b**, Central part of the nebula is shown in colour scale with yellow showing brightest parts and blue the faint emission. The structure of the bright optical jet is shown with black contours. Two extra contours are drawn for CO emission, at 12% and 20% of the peak intensity.

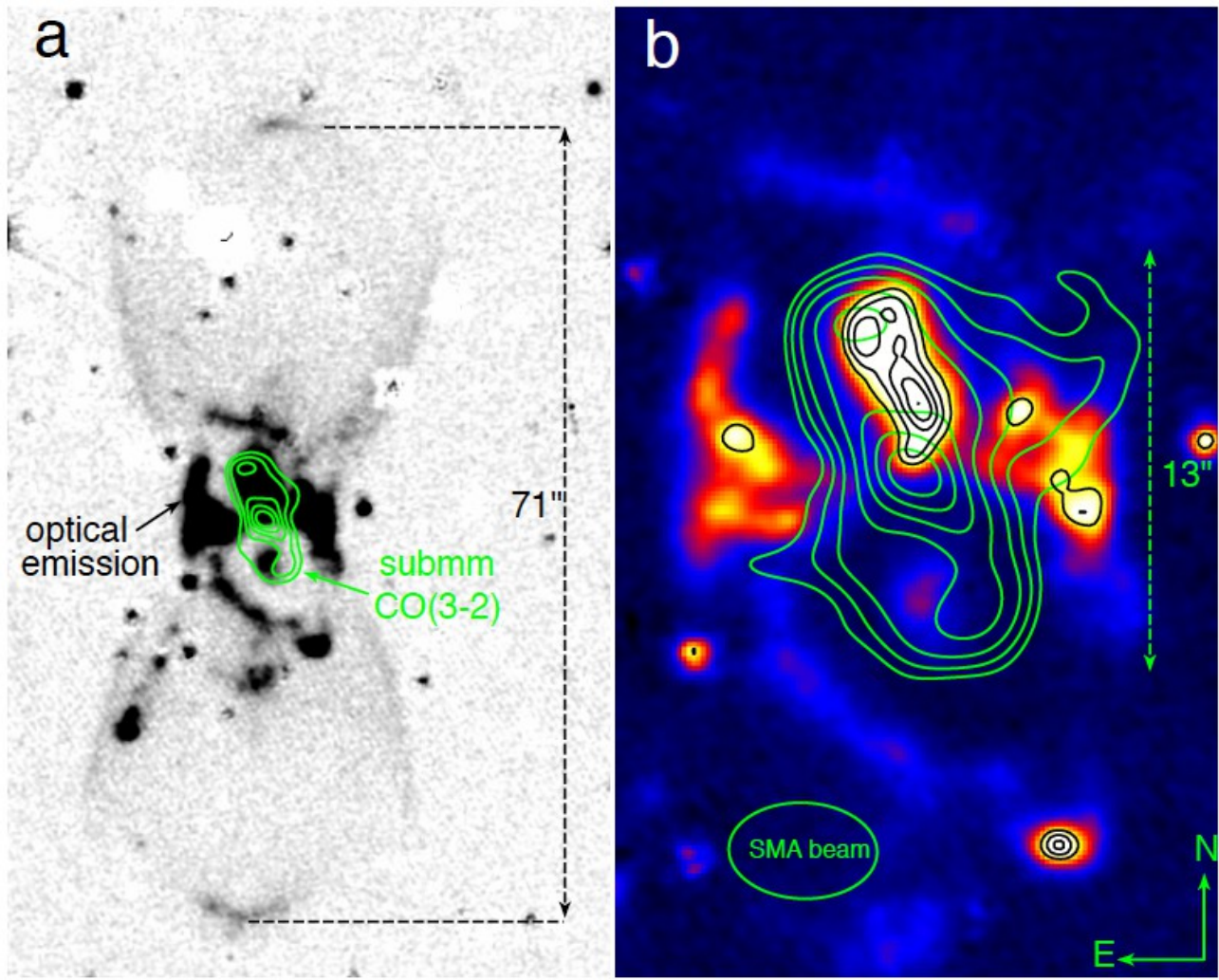

**Table 1 | Isotopic ratios of the molecular gas of CK Vul.**

| Isotopologue | Column-density ratio |
|---|---|
| $^{12}C^{16}O/^{13}C^{16}O$ | $6\pm2$ |
| $^{12}C^{16}O/^{12}C^{18}O$ | $23\pm15$ |
| $^{12}C^{16}O/^{12}C^{17}O$ | $\gg225$ |
| $H^{12}C^{14}N/H^{13}C^{14}N$ | $3\pm1$ |
| $H^{12}C^{14}N/H^{12}C^{15}N$ | $26\pm9$ |
| $^{12}C^{14}N/^{13}C^{14}N$ | $\sim2$ |
| $^{12}C^{14}N/^{12}C^{15}N$ | $\sim4^{\dagger}$ |
| $H^{12}CO^+/H^{13}CO^+$ | $2\pm1$ |
| $^{28}SiO/^{19}SiO$ | $4\pm4$ |

$^{\dagger}$based on uncertain identification

**METHODS**

**APEX observations.** CK Vul was observed with the Atacama Pathfinder Experiment (APEX) 12-m telescope[28] on several nights between 4 and 19 May 2014, and between 9 and 21 July 2014. Numerous frequency setups were observed between 217 and 909 GHz, all of which are listed in Extended Data Table 2. For observations up to 270 GHz, we used the SHeFI/APEX-1 receiver[29] which operates in a single sideband mode and produces spectra in a 4-GHz wide band. For frequencies between 278 and 492 GHz, we used the FLASH$^+$ receiver[30] which operates simultaneously in two atmospheric bands at about 345 and 460 GHz. Additionally, FLASH$^+$ separates the two heterodyne sidebands in the two 345/460 channels, giving four spectra simultaneously, each 4 GHz wide. Both APEX-1 and FLASH$^+$ are single-receptor receivers allowing for observation of one position at a time. For three of our setups with frequencies above 690 GHz, we used the CHAMP$^+$ receiver which consists of two arrays operating in the atmospheric windows at 660 and 850 GHz. Each CHAMP$^+$ array has seven receptors[31]. Each of the fourteen receptors of the CHAMP$^+$ array produced a single-sideband spectrum covering 2.8 GHz. As the backend for the APEX-1 and FLASH$^+$ observations, we used the eXtended Fast Fourier Transform Spectrometer[32] (XFFTS) which provided us with a spectral resolution of 88.5 kHz. The CHAMP$^+$ spectra were acquired with the array version of FFTS which operates at the spectral resolution of 732 kHz.

For most of our spectral setups which cover CO transitions up to $J$=4–3, we applied the position switching method with a reference at an offset (–180″,–100″) from CK Vul which was free of interstellar emission. Higher-$J$ transitions of CO, all lines of $^{13}$CO, and all setups which do not contain CO lines were observed with symmetric wobbler switching with a typical throw of 100″.

Observations were performed in weather conditions that were excellent or optimal for the given frequency setup. The typical system temperatures ($T_{sys}$) and rms noise levels reached are given in Extended Data Table 2 (the rms is specified for spectral binning given in the sixth columns of the table). The beam sizes and the main-beam efficiencies of the APEX antenna at each observed frequency is also given in the table. The typical calibration uncertainties are below 20%.

All spectra were reduced using standard procedures in the CLASS/GILDAS package.

**Effelsberg observations.** The Effelsberg 100-m Telescope was used to observe the classical circumstellar radio transitions: SiO(1–0) at $v=1$ and 2; four ground-state transitions of OH $^2\Pi_{3/2}$ (between 1.6 and 1.7 GHz); the $6_{1,6}$–$5_{2,3}$ transition of water at 22.235 GHz; and three lowest inversion lines of $NH_3$. From those, only the ammonia lines were detected and these observations are described in more detail below.

The three inversion lines of ammonia, $(J,K)=(1,1)$, (2,2), and (3,3) (the first two are *para* and the last is an *ortho* transition) we observed simultaneously on 2 August 2014. The secondary-focus receiver S13mm and the Effelsberg XFFTS were used. Spectra were centred at 23.750 GHz and covered 0.5 GHz at a resolution of 0.2 km/s. The spectra were moderately affected by baseline irregularities. The three lines of $NH_3$ are detected at a high signal-to-noise ratio (>10 for peaks), but baseline imperfections cast doubts on the actual profile and total intensity of the (3,3) line. The integration resulted in an rms noise level of 7.0 mK (in $T_{mb}$ scale) per 9 km/s bin. The telescope beam had a full width at half-maximum of 36.5".

Observations were repeated with the same instrumentation on 11 September 2014 but with the band centre shifted to lower frequencies to cover the (5,5) transition of $^{15}NH_3$ at 23.42 GHz. The spectra covered the (1,1) and (2,2) lines of $NH_3$, but not the (3,3) transition. At the rms of 4.1 mK ($T_{mb}$) per 10 km/s the line of $^{15}NH_3$ was not detected. This transition arises from a high level above the ground ($E_u$=296 K) and may be very weak in this source.

**SMA observations.** To image the emission of selected lines discovered with APEX at a higher angular resolution, we used the Submillimeter Array (SMA) on 3 and 30 July 2014. On 3 July 2014, the array was used in its compact configuration and with eight operating antennas. The phase centre for all the SMA observations of CK Vul was the position of the radio continuum source measured by the Very Large Array[3] (VLA), i.e. at RA=19:47:38.074 and Dec=+27:18:45.16. As absolute-flux calibrators, MWC349a and Uranus were observed, while 3C279 and 3C454.3 were observed for a bandpass calibration; 2025+337 and 2015+371 were our gain calibrators. The data covered four frequency ranges: 330.2–332.2, 335.2–337.2, 345.2–347.2, and 350.2–352.2 GHz. Although mainly aimed to observe the CO(3–2) transition, this setup gave us access to several emission lines and provided a very sensitive measurement of continuum emission. The system temperatures were changing between 200 and 500 K with the changing source elevation. The synthesized beam of these observations has an FWHM of 2.3"×1.5" and a position angle (PA) of 87°.6, while the primary beam, defining the field of view of the array, has an FWHM of 32".

On 30 July 2014, seven antennas were used in the subcompact configuration. The bandpass calibration was performed using observations of 3C279 while flux calibration was obtained by observing Mars and Titan. Gain calibrators were the same as earlier. The typical system temperatures were between 90 and 130 K. We covered four frequency ranges: 216.9–218.8, 218.9–220.8, 228.9–230.8, and 230.9–232.8 GHz. The synthesized beam of these observations was 8.4"×4.7" (PA=71°.5), while the primary beam at the observed frequencies is of 49".

The data were processed and calibrated in the MIR-IDL package. The calibrated visibilities were then imaged and further processed with Miriad[33]. The continuum emission was subtracted from the spectra as a best-fit first-order polynomial and continuum images were created by combining all four bands covering in total 8 GHz on each date. Resulting continuum flux densities are given in Extended Data Table 4.

A data inspection revealed that the interferometric maps of CO(3–2) show significantly lower flux than expected from the APEX spectra owing to the lack of short baselines. In the 345 GHz observation obtained in the compact configuration, the projected baselines gave us access to angular scales smaller than about 14". Any more extended emission was spatially filtered out by the interferometer. We corrected the interferometric observations by providing an APEX map covering a large part of the interferometer's field of view, i.e. 11"×11", and at a signal-to-noise ratio similar to that measured in the interferometric map. The two data sets were combined in Miriad using the *immerge* task.

**Identification of spectral features.** In the spectral survey obtained with APEX, we have identified 47 features to which we ascribed molecular transitions; three extra transitions were observed with the Effelsberg Telescope. All lines are listed in Extended Data Table 2. Ten of these features are very weak so that their presence and/or identification is uncertain. In the identification procedure, we referred to the Jet Propulsion Laboratory catalogue[34] and Cologne Database for Molecular Spectroscopy[35,36] (CDMS). Extended Data Table 2 includes basic measurements for the strongest features: the centroid position with respect to the laboratory frequency of the ascribed transition; the line FWHM in velocity units; and profile-integrated intensity of the line in $T_{mb}$ units. The list of detected transition includes mostly simple two-atomic species, but two molecules containing four atoms, i.e. $H_2CO$ and $NH_3$, were observed. Transitions of molecules containing H and CNO elements dominate the spectrum; those include CO, CN, HCN, HNC, $HCO^+$, $N_2H^+$, $H_2CO$ (and their isotopologues). The strongest are lines of carbon monoxide. Our survey covered four transitions of the main CO species and at least three transitions of its rare isotopologues. Only two unambiguously identified molecules are carriers of heavier atoms, i.e. SiO and CS, the latter being identified only tentatively. Two ionic species have been firmly identified, $HCO^+$ and $N_2H^+$. The most striking feature of the list of detected transitions is the high number of lines from rare isotopologues of CNO elements.

**Determination of abundances and excitation temperatures.** A few molecules were observed in multiple transitions within a range of upper-level energies ($E_u$) wide enough to allow a simple excitation analysis. With the aim to constrain the excitation temperatures and column densities, we performed analysis of rotational diagrams[37] (RDs), in which we assumed the thermodynamic equilibrium, optically thin emission, and that the gas is isothermal. Although some of the observed transitions are likely to be optically thick and the gas is not isothermal, this initial analysis was aimed to get the first constraints on the gas physical parameters. We used least-square fitting to derive the physical parameters. Partition functions were interpolated from data tabulated in CDMS. The sizes of the emission regions, necessary for a beam-filling correction, were based on our interferometric maps.

Our RD analysis was supported by spectra simulations performed in CASSIS[38]. The tool allowed us to generate a model spectrum with line profiles approximated by Gaussians. The simulation was based on the same assumptions as underlying the RD analysis, but included a limited correction for line saturation effects. The CASSIS simulation was especially helpful in an analysis of blended features and transitions with considerable hyperfine splitting, for instance CN and its isotopologues.

Rotational diagrams for CO and $H^{13}CN$ which were observed also in transitions with $E_u>80$ K, cannot be reproduced by a simple linear fit. This is likely a consequence of multiple gas components at different temperatures (or a continues range of temperatures), combined with different sizes of the emission regions contributing most to the given transition. Additionally, those transitions at high $E_u$ were typically observed at high frequencies at which the APEX beam is significantly smaller than for the rest of observed transitions and does not encompass the entire molecular region. Because of the missing spatial information, those transitions were omitted in the RD analysis. Here we focus on the gas at lower temperatures which dominates the emission in lower rotational transitions.

For most molecules analysed here, the excitation temperature was derived from the RD of the isotopologue for which the highest number of transitions was observed. Then, the same temperature was assumed for other isotopologues, and column densities were calculated for all other isotopic species observed in at least one transition. For all three CO isotopologues, good temperature estimates were obtained for each isotopologue and the final column densities were calculated for a weighted mean of the three values. While the relative abundances of the different species analysed here are subject to large errors (mainly because of the complex spatio-kinematical structure of the gas), the isotopic ratios are much more reliable — they weakly depend on the temperature and are not directly sensitive to the details of the spatial distribution (if no chemical fractionation takes place). They are, however, affected by opacity effects (see below). In Table 1, we therefore report only the isotopic ratios resulting from our analysis. To put constrains on species containing the oxygen isotope $^{17}$O, we used the upper limit on the C$^{17}$O(3–2) line covered by APEX.

The $^{12}$C/$^{13}$C ratio was derived for four species and is consistently found to be 2–3 for three of them. The value derived from CO line ratios is an outlier, with slightly *higher* ratio of about 6. The saturation effect, if present, should be strongest in the CO transitions giving a ratio which is *lower* than in the weaker lines of the rarer species. The nitrogen-bearing species, HCN and CN, lead to two different values of the isotopic $^{14}$N/$^{15}$N ratio, 26 and 4, respectively. There is an extra uncertainty in the abundance analysis of the weak C$^{15}$N spectra related to their hyperfine structure and blending. Chemical fractionation cannot be excluded because CN is likely a product of photodissociation of HCN and self-shielding effects are likely to occur for HCN isotopologues.

The RD analysis allowed us to derive excitation temperatures for the different species. They are typically in the range 8-22 K, but – as noted above – extra gas components at higher temperatures are evident in transitions from higher energy levels.

We tried to assess the influence of the line saturation effects on the results of our analysis by investigating the optical depth of the CO lines, which are expected to have the highest opacity. We analysed the CO emission over the full line profile ($V_{\rm LSR}$=–220 to 200 km/s) and also in one wing (–50 to 40 km/s). The opacity was calculated for the best-fit parameters of temperature and column density. The line FWHM was set to 120 and 90 km/s for the full profile and the probed part of the wing, respectively. For a source size of 10", whose solid angle is equivalent to that of the entire emission region seen in the combined SMA and APEX maps, we get an optical thickness of $\tau_0(N_{\rm u})$=0.35 for CO(2–1) (strongest feature observed) and lower values for the weaker lines. For the emission in the wing, we get $\tau_0(N_{\rm u})$=0.18 for the $J$=2–1 transition and much less for the higher-$J$ lines. However, the obtained results are sensitive to the adopted value of the source size. For FWHM=6.8", which corresponds to size of the CO(3–2) emission at the isophote at the 30% of the peak, the strongest line would have an optical thickness of 0.75. Then the central CO component, which is of an even smaller size of 1"–2" and contributes about 15% of the total observed flux, produces emission of moderate opacity of the order of 1. Only if the emission arises in compact clumps, lines are optically thick.

**APEX observations of other Galactic novae.** Our detection of CO in CK Vul contradicts the previous claims of non-detection of rotational circumstellar lines in this source[39]. It also casts doubts on all earlier negative results of searches of submillimetre-wave lines towards novae and related objects. Observation of lines as broad as those expected in novae (300–7,000 km/s) are very demanding in terms of the atmospheric and instrumental stability. In the earlier attempts, lines were often broader than the full available spectral range of the receiving systems or comparable in width to typical baseline ripples. The presence of molecular emission in classical novae was therefore tested anew using the modern instrumentation of APEX.

The novae observed with APEX were selected from *CBAT List of Novae in the Milky Way*[40] using the following selection criteria: (1) the source has to reach elevations higher than 40°, and (2) has to be available for observations in the LST range 23–13 hours to not collide with the inner-Galaxy projects in the APEX observing queue, (3) must be located at least 3° from the Galactic plane to avoid contamination from Galactic CO emission. These requirements limited the number of sources to 17, which are listed in Extended Data Table 2.

The observations were performed between 24 and 28 August 2014 and on 8 September 2014 with FLASH$^+$ connected to FFTS providing a spectral coverage of 4 GHz. Although four spectral ranges were covered simultaneously, the observing procedure was optimised for the band centred at the frequency of the CO(3–2) line. The observations were performed with wobbler switching with a throw of 80". No source was detected in the CO(3–2) line at the typical rms of 2.5 mK ($T_{mb}$) per 33 km/s (Extended Data Table 3). At the same sensitivity the line was very clearly seen in the spectrum of CK Vul.

**Spectral energy distribution.** Using archival and literature data combined with our SMA continuum measurements, we constructed the spectral energy distribution (SED) of CK Vul. The data are described in detail at the end of this section; the measurements are summarised in Extended Data Table 4 and shown in Extended Data Fig. 1.

The SED is dominated by emission ranging from about 20 μm up to the millimetre wavelengths. The flux density $F_\nu$ peaks at about 100 μm. The long-wavelength part of the $F_\nu$ distribution, from the far-infrared (FIR) to the SMA measurement, has a slope with a spectral index α=2.1±0.1 (where $F_\nu \propto \nu^\alpha$) and can be interpreted as thermal dust radiation. A single black-body cannot explain the observed emission entirely but the best fit of a single Planck function provides a rough estimate of the dust temperature of 39±5 K. The best fit of a grey-body, i.e. a Planck function multiplied by dust emissivity in the form of a power law $\nu^\beta$, gives a temperature of 15 K and β=1.0. This fit underestimates the source fluxes at shorter wavelengths, but we believe it provides a good estimate on the value of β. Moreover, β≅1.0 is expected for circumstellar dust in the form of amorphous carbon or layer-lattice silicates[41]; β≈1.0 is also typical for circumstellar disks[42,43]. We note that the chemical composition and the form (crystalline/amorphous) of dust in CK Vul remains completely unknown. In order to better reproduce the flux at short wavelengths, we also obtained a fit of two grey bodies with β being fixed at a value of 1.0. This gave temperatures of 15 and 49 K. It is unlikely that the dust is characterized by two isothermal components. Instead, one can expect a continuous range of temperatures in 15–49 K. The fit of two grey bodies (Extended Data Fig. 1) underestimates the fluxes around 160 μm. Although this could be overcome by introducing an extra component at an intermediate temperature, we did not attempted it because the least-square fits become degenerate at the required number of parameters.

The flux under the reconstructed SED is $6.0 \cdot 10^{-11}$ erg/s/cm². Adopting the distance of 700 pc (ref. 3), we calculate the source luminosity of $3.6 \cdot 10^{33}$ erg/s (or 0.9 $L_\odot$). This luminosity is close to 0.7 $L_\odot$ found from ionization-equilibrium calculations for the optical nebula[3] (here corrected to the distance of 700 pc). The dust emission we observe must be reprocessed radiation of the central source which is hidden for our line of sight at wavelengths shorter than ~20 μm. Because the obscuring material has a form of a flattened, torus-like structure, the radiation field within the whole system is anisotropic. Our estimate should therefore be treated as a lower limit on the actual luminosity of the source.

The continuum observations and data reductions were the following.

*Herschel*: On 23 October 2011, CK Vul was serendipitously observed by photometers on board the Herschel Space Observatory in a field covered within the Hi-Gal project[44]. Two scans (OBSIDs 1342231339 and 1342231340) were obtained in orthogonal directions across a large field covering CK Vul. The two *Herschel* cameras, PACS and SPIRE, were used simultaneously in these observations. In both scans, PACS was used with its blue (70 μm) and red (160 μm) bands (i.e. the green band was not used) and SPIRE produced maps in its all three bands, i.e. 250, 350, and 500 μm. Data were retrieved from the Herschel Science Archive and processed in Herschel Interactive Processing Environment (HIPE). The raw data were automatically reduced by the standard pipeline which used the calibration scheme version 12.1. Pointing accuracy of *Herschel* is typically 2" and the source we identify as CK Vul has a position which is consistent within 3" with the position of the continuum seen by VLA and SMA. In all the observed bands, CK Vul appears as a point source, but its background becomes more and more contaminated by diffuse Galactic emission with increasing wavelength. A bright source closest to CK Vul is located 1.5 arcmin west. It is weaker than CK Vul in all the PACS and SPIRE bands.

Source fluxes in the four individual PACS maps were measured with aperture-photometry techniques including background subtraction and a correction for limited aperture size. Results obtained for the two PACS bands were averaged and the standard deviation from the two measurements in each band was taken as the uncertainty.

Source fluxes in the SPIRE observations were measured using aperture photometry tasks and a 'timeline fitting' procedure available in HIPE. In addition to an aperture correction, we also applied a colour correction to the measured fluxes using tabular data included in the *Herschel*-SPIRE calibration data for the spectral index of $\alpha=1.0$. Measurements were obtained on individual scans and the results were averaged for the given band. The uncertainties in the absolute flux calibration are of 6% for SPIRE, and 10% and 20% for the blue and red bands of PACS, respectively.

*Spitzer*: CK Vul was observed multiple times with *Spitzer* instruments. The Multiband Imaging photometer for Spitzer (MIPS) operating in bands at 24, 70, and 160 μm observed the position on two different dates, i.e. on 17 October 2004 a MIPS scan centred on the object was obtained (AOR 10837504, PI A. Evans) and on 7 October 2005 the position was covered by a scan aimed to observe Galactic emission in the field of CK Vul (AOR 15621888, PI S. Carey; no data in the 160 μm band were collected). We used the pipeline processed data and aperture-photometry procedures to derive the source fluxes. The aperture- and colour-corrected (for the assumed black body spectrum of 30 K) fluxes are listed in Extended Data Table 4. For the 24 and 70 μm bands we list the average flux from the two scans and the standard deviation from the mean as an error. The single observation in the 160 μm was spatially under-sampled and only a very rough flux estimate was performed. The flux is indeed lower than the PACS measurement at similar wavelengths and was omitted in the analysis. The measurement at 70 μm, on the other hand, agrees very well with that from PACS at a similar wavelength. At the angular resolution of the MPIS maps at 24 and 70 μm of 6″ and 18″ (FWHM), respectively, the source appears point-like.

The InfraRed Array Camera (IRAC) observed the positions of CK Vul four times in October 2004 and December 2012 with a different combination of IRAC bands (3.6, 5.8, 4.5, and 8.0 μm). None of the IRAC maps shows a measurable source at the position of CK Vul. We used the most sensitive scans in the 3.6 and 8.0 μm to derive upper limits on the emission from CK Vul. The standard deviation of the flux at the position of the object is of about $\sigma=1.21$ μJy and 4.24 μJy in the 3.6 and 8.0 μm bands, respectively.

*WISE*: The point source catalogue of the Wide-field Infrared Survey Explorer (WISE) survey lists a source consistent with the position of CK Vul which was measured in three out of the four WISE bands (there is only an upper limit in the *W3* band). The source catalogue position is about 5″ away from the SMA position of the continuum source. The catalogue flags also indicate that the source is resolved (FWHM of the point-spread functions are 6.1, 6.4, 6.5, and 12 arcseconds in the *W1* to *W4* bands.) The flags also indicate the source is variable in the *W1* and *W2* bands. After inspecting the WISE images covering the position of CK Vul and comparing them to optical and radio maps, we concluded that only the *W4* measurement at 22 μm can be undoubtedly ascribed to the source seen at longer wavelengths (while the *W1* and *W2* data correspond to 'variable 2' identified in a recent study[16]). The average magnitudes from the WISE point source catalogue (resulting from profile fitting) were converted to flux units using standard zero points[45] and are listed in Extended Data Table 4. The catalogue values in *W1* to *W3* bands can all be treated here as rough upper limits on the flux of CK Vul. No colour correction was applied.

*AKARI*: AKARI's point source catalogues[46] contain one source which matches the positions of CK Vul. The measurements obtained with the Far-Infrared Surveyor (FIS) instrument which operates at 65, 90, 140, and 160 μm are flagged as reliable only for the measurement at 90 μm (722 mJy). In the source catalogue of the Infrared Camera (IRC) survey at 9 and 18 μm, no source can be identified as CK Vul.

*JCMT:* Literature data[47] exist based on observations obtained with the James Clerk Maxwell Telescope (JCMT) and the SCUBA bolometer at about 450 and 850 μm (~667 and ~353 GHz). The measurement at 850 μm covers a wavelength range close to that of one of our SMA observations. The SCUBA flux is slightly above that derived in the SMA observations. The reason for this is likely the fact that our SMA measurements represent only line-free continuum while the bolometric observations represent summary flux of continuum and emission lines. Our APEX spectra in the range between 333 and 357 GHz, which overlap with a high-sensitivity part of the SCUBA-850 μm bandpass, show a line flux density of 94.3 mJy, what constitutes 42% of the flux measured with SCUBA. Spectral lines contribute therefore significantly to the bolometric measurements, at least in the submillimetre-wave region. To a lesser degree, the SMA continuum measurement at 341 GHz can be partially lower than that measured with JCMT because extended continuum emission, if present, was partially filtered out by the interferometer.

The SCUBA measurement at 450 μm of is close in wavelength to the SPIRE 500 μm band (482.3 μm), but has a significantly lower flux. Compared to all the data collected, this SCUBA measurement is a clear outlier. Because the ground-based observations at 450 μm are very demanding in terms of weather conditions, we suspect this measurement has an extra systematic uncertainty not quoted in the work reporting the data[47].

On 3 August 2012 CK Vul was observed again with the JCMT, this time with the SCUBA-2 bolometer array. While no source was detected at 450 μm, the emission at 850 μm is very clear. Using the archival pipeline-processed data, we measured the source flux to be 194.0±1.7 mJy (1σ error). This measurement agrees within the uncertainties with the flux measured in observations taken eleven years earlier[47]. The source is unresolved at the resolution of 14.5 arcseconds (FWHM). The 3σ upper limit on the flux density at 450 μm is of 1.08 Jy.

*VLA and Toruń*: For completeness, in the SED analysis we include flux measurements of the radio continuum obtained with the VLA[3,48] and Toruń Radiotelescope[49].

**Table ED1 | List of detected transitions with Effelsberg and APEX telescopes.** Notes: **a)** Lines detected in the SMA observations are indicated with a check mark. **b)** Centroid of the line with respect to the local standard of rest and the 1σ error. **c)** Full width at half-maximum and 1σ error resulting from a Gaussian fit. **d)** Intensity of the line or 3σ upper limit. The uncertainties given are related to the spectral 1σ noise level. **e)** Flux and width measurements uncertain owing to baseline irregularity. **f)** Weakly contaminated by $^{13}$CN. **g)** Central laboratory frequency estimated as a mean of multiple hyperfine components weighted by their respective intensities ($A_{ul}$). Summary $A_{ul}$ is given. **h)** Line blends with the feature of $C^{18}O$. **i)** Detection at a low statistical significance or doubtful. **j)** Weakly contaminated by the line of $^{15}$CN. **k)** Integrated-intensity measurement include contribution form the interstellar features. **l)** Uncertain identification.

| Molecule | Transition | Frequency lab. (MHz) | $E_u$ (K) | $A_{ul}$ (s$^{-1}$) | Detect. SMA[a] | $V_{LSR}$[b] (km s$^{-1}$) | FWHM[c] (km s$^{-1}$) | $\int T_{mb}\,dv$[d] (K km s$^{-1}$) | Notes |
|---|---|---|---|---|---|---|---|---|---|
| NH$_3$ | $J,K = 1,1$ para | 23694.50 | 23.3 | 1.68E−7 | | −9.2±1.8 | 99.2±1.9 | 23.70±0.25 | |
| NH$_3$ | $J,K = 2,2$ para | 23722.63 | 64.4 | 2.24E−7 | | −14.5±3.4 | 89.3±3.9 | 6.50±0.23 | |
| NH$_3$ | $J,K = 3,3$ ortho | 23870.13 | 123.5 | 2.57E−7 | | −26.6±3.4 | 86.1±3.9 | 6.35±0.29 | e |
| SiO | $J = 5−4$ | 217104.98 | 31.3 | 5.20E−4 | ✓ | −16.0±2.8 | 78.8±3.0 | 4.29±0.14 | f |
| $^{13}$CN | $N = 2−1, J = 3/2−1/2$ | 217297.72 | 15.7 | 5.23E−4 | ✓ | 8.2±7.1 | 79.3±9.7 | 1.73±0.12 | g |
| $^{13}$CN | $N = 2−1, J = 5/2−3/2$ | 217456.59 | 15.7 | 5.31E−4 | ✓ | −16.5±11.7 | 80.8±15.8 | 1.47±0.11 | g |
| $^{13}$CN | $N = 2−1, J = 3/2−3/2$ | 217633.04 | 15.7 | 1.14E−5 | | 12.3±7.0 | 33.1±7.0 | ≤0.26 | g |
| H$_2$CO | $J_{K_a,K_c} = 3_{0,3} − 2_{0,2}$ | 218222.19 | 21.0 | 2.82E−4 | ✓ | −0.7±17.4 | 25.0±21.7 | ≤0.28 | |
| H$_2$CO | $J_{K_a,K_c} = 3_{2,2} − 2_{2,1}$ | 218475.63 | 68.1 | 1.57E−4 | ✓ | 25.4±9.0 | 21.7±10.8 | ≤0.24 | |
| C$^{15}$N? | $N = 2−1, J = 3/2−3/2$ | 219026.81 | 15.8 | 3.46E−5 | ✓ | −41.3±8.6 | 34.7±8.8 | ≤0.24 | g h i |
| C$^{18}$O | $J = 2−1$ | 219560.35 | 15.8 | 6.01E−7 | ✓ | −5.6±4.2 | 90.7±4.6 | 2.05±0.11 | j |
| C$^{15}$N? | $N = 2−1, J = 3/2−1/2$ | 219722.80 | 15.8 | 1.73E−4 | ✓ | −79.3±2.0 | 1.2±2.0 | ≤0.17 | i |
| C$^{15}$N | $N = 2−1, J = 5/2−3/2$ | 219933.63 | 15.8 | 2.08E−4 | ✓ | −24.4±11.5 | 53.4±13.4 | 0.60±0.11 | g |
| $^{13}$CO | $J = 2−1$ | 220398.68 | 15.9 | 6.07E−7 | ✓ | −20.7±1.4 | 101.8±1.4 | 20.79±0.18 | |
| H$_2$CO | $J_{K_a,K_c} = 3_{1,2} − 2_{1,1}$ | 225697.78 | 33.5 | 2.77E−4 | | 19.8±9.0 | 39.8±9.1 | ≤0.57 | |
| CN | $N = 2−1, J = 3/2−1/2$ | 226658.92 | 16.3 | 2.85E−4 | | −15.9±6.6 | 52.2±7.2 | 2.01±0.22 | g |
| CN | $N = 2−1, J = 5/2−3/2$ | 226876.46 | 16.3 | 3.43E−4 | | −12.6±5.8 | 51.1±7.0 | 2.47±0.18 | g |
| CO | $J = 2−1$ | 230538.00 | 16.6 | 6.91E−7 | ✓ | −1.6±2.4 | 100.7±2.4 | 51.01±1.45 | k |
| $^{13}$CS? | $J = 5−4$ | 231220.69 | 33.3 | 2.51E−4 | ✓ | −62.0±5.9 | 98.9±6.2 | 1.31±0.19 | l |
| $^{29}$SiO? | $J = 6−5$ | 257255.22 | 43.2 | 8.78E−4 | | −9.2±7.5 | 75.1±8.1 | 1.01±0.08 | i l |
| HC$^{15}$N | $J = 3−2$ | 258157.00 | 24.8 | 7.65E−4 | | −44.4±8.0 | 69.2±8.6 | 0.46±0.12 | |
| H$^{13}$CN | $J = 3−2$ | 259011.80 | 24.9 | 7.72E−4 | | −11.2±1.3 | 74.3±1.4 | 8.46±0.15 | |
| H$^{13}$CO$^+$ | $J = 3−2$ | 260255.34 | 25.0 | 1.34E−3 | | −17.8±1.9 | 18.8±2.0 | 0.56±0.07 | |
| SiO | $J = 6−5$ | 260518.02 | 43.8 | 9.12E−4 | | −15.4±3.6 | 62.3±3.7 | 2.51±0.14 | |
| HCN | $J = 3−2$ | 265886.43 | 25.5 | 8.36E−4 | | −5.7±2.4 | 74.1±2.5 | 16.45±0.69 | |
| N$_2$H$^+$ | $J = 3−2$ | 279511.73 | 26.8 | 1.35E−3 | | −42.7±6.1 | 83.4±7.5 | 1.27±0.07 | |
| H$_2$CO | $J_{K_a,K_c} = 4_{1,4} − 3_{1,3}$ | 281526.93 | 45.6 | 5.88E−4 | | 17.8±13.5 | 83.9±21.2 | 0.49±0.05 | |
| CS | $J = 6−5$ | 293912.09 | 49.4 | 5.23E−4 | | −8.7±3.7 | 49.5±3.9 | 0.53±0.06 | |
| $^{29}$SiO? | $J = 7−6$ | 300120.48 | 57.6 | 1.41E−3 | | 6.9±8.5 | 45.6±10.5 | 0.30±0.05 | l |
| C$^{18}$O | $J = 3−2$ | 329330.55 | 31.6 | 2.17E−6 | | −16.1±17.1 | 76.1±18.0 | 1.68±0.38 | i |
| $^{13}$CO | $J = 3−2$ | 330587.97 | 31.7 | 2.19E−6 | ✓ | −10.0±2.0 | 97.9±2.0 | 20.40±0.51 | |
| CN | $N = 3−2, J = 5/2−5/2$ | 339487.80 | 32.6 | 8.18E−5 | | −43.0±12.3 | 24.2±13.1 | ≤0.28 | g i |
| CN | $N = 3−2, J = 5/2−3/2$ | 340031.29 | 32.6 | 1.15E−3 | | −0.5±5.2 | 105.6±6.0 | 3.26±0.16 | g |
| CN | $N = 3−2, J = 7/2−5/2$ | 340248.80 | 32.7 | 1.24E−3 | | −18.4±5.0 | 81.3±5.7 | 3.26±0.17 | g |
| CS | $J = 7−6$ | 342882.85 | 65.8 | 8.40E−4 | | −52.5±8.4 | 65.0±9.3 | ≤0.48 | |
| $^{29}$SiO? | $J = 8−7$ | 342980.84 | 74.1 | 2.12E−3 | | 33.5±0.3 | 62.0±0.3 | ≤0.30 | i l |
| HC$^{15}$N | $J = 4−3$ | 344200.11 | 41.3 | 1.88E−3 | | 12.4±7.1 | 24.4±7.5 | ≤0.25 | i |
| H$^{13}$CN | $J = 4−3$ | 345339.77 | 41.4 | 1.90E−3 | ✓ | −3.4±1.9 | 59.0±1.9 | 2.64±0.13 | |
| CO | $J = 3−2$ | 345795.99 | 33.2 | 2.50E−6 | ✓ | −13.1±0.7 | 92.9±0.7 | 47.03±0.34 | |
| H$^{13}$CO$^+$ | $J = 4−3$ | 346998.34 | 41.6 | 3.29E−3 | ✓ | −5.8±4.5 | 39.9±4.8 | 0.69±0.19 | |
| SiO | $J = 8−7$ | 347330.58 | 75.0 | 2.20E−3 | | −13.2±6.2 | 87.5±7.2 | 1.76±0.13 | |
| HN$^{13}$C | $J = 4−3$ | 348340.90 | 41.8 | 2.03E−3 | | 4.0±7.1 | 30.7±7.3 | 0.34±0.09 | |
| CCH? | $N = 4−3, J = 9/2−7/2, 7/2−5/2$ | 349364.58 | 41.9 | 7.26E−4 | | 0.1±34.7 | 38.8±63.0 | ≤0.30 | i |
| HCN | $J = 4−3$ | 354505.48 | 42.5 | 2.05E−3 | | −8.7±1.2 | 77.0±1.3 | 8.98±0.16 | |
| HCO$^+$ | $J = 4−3$ | 356734.22 | 42.8 | 3.57E−3 | | −17.1±5.8 | 27.2±6.2 | 0.71±0.11 | |
| CS | $J = 8−7$ | 391846.89 | 84.6 | 1.26E−3 | | 1.4±0.3 | 45.6±0.3 | 0.66±0.14 | i |
| HC$^{15}$N | $J = 5−4$ | 430235.32 | 62.0 | 3.75E−3 | | −22.2±2.8 | 26.3±2.8 | 1.98±0.29 | l |
| CO | $J = 4−3$ | 461040.77 | 55.3 | 6.13E−6 | | −22.7±2.1 | 89.2±2.2 | 28.95±0.73 | |
| H$^{13}$CN | $J = 8−7$ | 690552.08 | 149.2 | 1.61E−2 | | −36.9±17.7 | 52.0±21.4 | 2.72±0.72 | |
| CO | $J = 6−5$ | 691473.08 | 116.2 | 2.14E−5 | | −68.3±6.6 | 110.2±7.1 | 24.87±1.12 | |

**Table ED2 | Spectral setups observed with the APEX telescope.** Notes: **a )** Spectral range overlaps with the next setup in the table. **b)** Part of the integration obtained also on 08-05-2014. **c)** Part of the integration obtained also on 06-05-2014.

| Central freq. (MHz) | $T_{sys}$ (K) | Time ON (min) | Bandwidth (GHz) | rms $T_A^*$ (mK) | rms-bin (km s$^{-1}$) | $\eta_{mb}$ | Beam FWHM (″) | Date 2014 DD-MM |
|---|---|---|---|---|---|---|---|---|
| 218,800.0 | 151 | 39.7 | 4.0 | 5.9 | 1.05 | 0.75 | 28.5 | 09-05 |
| 226,739.6 | 183 | 15.1 | 4.0 | 11.5 | 1.01 | 0.75 | 27.5 | 05-05 |
| 230,038.0 | 165 | 0.6 | 4.0 | 50.0 | 0.99 | 0.75 | 27.1 | 04-05 |
| 230,538.0 | 183 | 12.8 | 4.0 | 13.8 | 0.99 | 0.75 | 27.1 | 05-05 |
| 259,011.8 | 258 | 93.5 | 4.0 | 6.8 | 0.88 | 0.73 | 24.1 | 10-05 |
| 266,721.9 | 268 | 5.5 | 4.0 | 31.2 | 0.86 | 0.73 | 23.4 | 05-05 |
| 280,500.0 | 156 | 53.4 | 4.0 | 3.8 | 1.22 | 0.72 | 22.2 | 09-07$^c$ |
| 287,364.0$^a$ | 162 | 26.1 | 4.0 | 6.0 | 1.19 | 0.72 | 21.7 | 19-05 |
| 288,507.1 | 164 | 4.1 | 4.0 | 14.5 | 1.19 | 0.72 | 21.6 | 06-05 |
| 292,498.5 | 155 | 53.4 | 4.0 | 4.0 | 1.17 | 0.72 | 21.3 | 09-07$^c$ |
| 299,362.5$^a$ | 154 | 26.1 | 4.0 | 5.4 | 1.15 | 0.72 | 20.8 | 19-05 |
| 300,505.6 | 169 | 4.1 | 4.0 | 16.1 | 1.14 | 0.71 | 20.8 | 06-05 |
| 313,154.4 | 175 | 2.7 | 4.0 | 19.1 | 1.10 | 0.71 | 19.9 | 06-05 |
| 325,152.9 | 2,144 | 2.7 | 4.0 | 267.0 | 1.06 | 0.70 | 19.2 | 06-05 |
| 330,588.0$^a$ | 393 | 21.9 | 4.0 | 17.0 | 1.04 | 0.70 | 18.9 | 08-05 |
| 333,797.5$^a$ | 240 | 14.0 | 4.0 | 10.6 | 1.37 | 0.70 | 18.7 | 06-05 |
| 336,301.5$^a$ | 186 | 40.0 | 4.0 | 4.8 | 1.36 | 0.70 | 18.6 | 14-05 |
| 340,247.8$^a$ | 192 | 22.4 | 4.0 | 6.5 | 1.34 | 0.70 | 18.3 | 09-07$^c$ |
| 342,586.4 | 266 | 21.9 | 4.0 | 9.2 | 1.34 | 0.69 | 18.2 | 08-05 |
| 343,601.5$^a$ | 196 | 44.1 | 4.0 | 4.8 | 1.33 | 0.69 | 18.2 | 15-05 |
| 345,796.0$^a$ | 228 | 14.0 | 4.0 | 9.9 | 1.32 | 0.69 | 18.0 | 06-05 |
| 348,300.0$^a$ | 196 | 40.0 | 4.0 | 5.1 | 1.31 | 0.69 | 17.9 | 14-05 |
| 352,246.3$^a$ | 237 | 22.4 | 4.0 | 8.4 | 1.30 | 0.69 | 17.7 | 09-07$^c$ |
| 355,600.0 | 228 | 44.1 | 4.0 | 5.7 | 1.29 | 0.69 | 17.5 | 15-05 |
| 393,749.9 | 545 | 28.4 | 4.0 | 16.7 | 1.16 | 0.67 | 15.8 | 09-07 |
| 398,306.3 | 484 | 56.4 | 4.0 | 10.4 | 1.15 | 0.67 | 15.7 | 15-05 |
| 403,086.0$^a$ | 452 | 14.2 | 4.0 | 18.4 | 1.13 | 0.67 | 15.5 | 09-07 |
| 405,748.8$^a$ | 416 | 28.4 | 4.0 | 12.8 | 1.13 | 0.66 | 15.4 | 09-07 |
| 407,963.5$^a$ | 442 | 26.1 | 4.0 | 13.9 | 1.12 | 0.66 | 15.3 | 19-05 |
| 410,304.8 | 449 | 56.4 | 4.0 | 9.9 | 1.11 | 0.66 | 15.2 | 15-05 |
| 415,085.0 | 449 | 14.2 | 4.0 | 18.3 | 1.10 | 0.66 | 15.0 | 09-07 |
| 419,962.0 | 610 | 26.1 | 4.0 | 22.0 | 1.09 | 0.66 | 14.9 | 19-05 |
| 428,766.7 | 1,165 | 40.4 | 4.0 | 38.0 | 1.07 | 0.65 | 14.6 | 08-05 |
| 440,765.2 | 5,137 | 40.4 | 4.0 | 167.8 | 1.04 | 0.65 | 14.2 | 08-05 |
| 461,040.8 | 1,085 | 48.0 | 4.0 | 25.4 | 0.99 | 0.64 | 13.5 | 08-05 |
| 473,039.2 | 4,949 | 48.0 | 4.0 | 158.1 | 1.45 | 0.63 | 13.2 | 08-05 |
| 479,201.5 | 1,186 | 7.9 | 4.0 | 59.8 | 1.43 | 0.63 | 13.0 | 06-05 |
| 491,200.0 | 1,659 | 7.9 | 4.0 | 91.5 | 1.40 | 0.62 | 12.7 | 06-05 |
| 691,473.1 | 2,728 | 88.3 | 2.8 | 16.2 | 3.18 | 0.53 | 9.0 | 17-05 |
| 806,651.8 | 6,236 | 8.2 | 2.8 | 121.8 | 2.72 | 0.47 | 7.7 | 17-05 |
| 909,158.8 | 17,593 | 80.1 | 2.8 | 117.0 | 2.41 | 0.42 | 6.9 | 17-05 |

**Table ED3 | Novae observed with APEX in search for CO(3-2) emission.** Upper limits on the emission are given in therms of the root-mean-squared (rms; in $T_{mb}$ scale per 33 km/s). Sources marked with an asterisk were observed in an earlier study[39] in CO(1-0).

| Source | RA | Dec | rms (mK) |
|---|---|---|---|
| CG CMa | 07:04:05.05 | -23:45:34.60 | 2.0 |
| CN Vel | 11:02:38.57 | -54:23:09.50 | 2.6 |
| CQ Vel | 08:58:50.99 | -53:20:17.80 | 2.2 |
| DY Pup | 08:13:48.51 | -26:33:56.50 | 2.0 |
| GQ Mus* | 11:52:02.35 | -67:12:20.20 | 3.2 |
| GU Mus | 11:26:26.60 | -68:40:32.30 | 2.8 |
| LZ Mus | 11:56:09.27 | -65:34:20.20 | 3.0 |
| RR Pic | 06:35:36.06 | -62:38:24.30 | 2.6 |
| T Pyx* | 09:04:41.50 | -32:22:47.50 | 2.2 |
| TV Crv | 12:20:24.15 | -18:27:02.00 | 2.3 |
| V1065 Cen | 11:43:10.33 | -58:04:04.30 | 3.9 |
| V351 Car | 10:45:19.14 | -72:03:56.00 | 3.4 |
| V359 Cen | 11:58:15.33 | -41:46:08.40 | 2.3 |
| V382 Vel | 10:44:48.39 | -52:25:30.70 | 1.7 |
| V598 Pup | 07:05:42.51 | -38:14:39.40 | 2.1 |
| VX For | 03:26:45.71 | -34:26:25.20 | 2.4 |
| WX Cet | 01:17:04.17 | -17:56:23.00 | 2.2 |

**Table ED4 | Continuum measurements of CK Vul used in the analysis of its spectral energy distribution.** Entries marked with an asterisk represent flux of pure continuum, i.e. with no contribution from spectral lines.

| Telescope-instrument-band | $\lambda$ ($\mu$m) | $F_\nu$ or $3\sigma$ upper limit (Jy) |
|---|---|---|
| WISE-$W1$ | 3.4 | $\ll 1.32E-3$ |
| WISE-$W2$ | 4.6 | $\ll 0.82E-3$ |
| WISE-$W3$ | 11.6 | $< 0.65E-3$ |
| WISE-$W4$ | 22.1 | $(16.50\pm1.11)E-3$ |
| *Spitzer*-IRAC-3.6 | 3.6 | $< 3.62E-6$ |
| *Spitzer*-IRAC-8.0 | 8.0 | $< 12.7E-6$ |
| *Spitzer*-MIPS-24 | 23.68 | $(9.64\pm0.29)E-3$ |
| *Spitzer*-MIPS-70 | 71.42 | $0.503\pm0.050$ |
| *Spitzer*-MIPS-160 | 155.90 | $0.996\pm0.100$ |
| AKARI-FIS-90 | 76.9 | $0.722\pm0.07$ |
| *Herschel*-PACS-70 | 68.9 | $0.484\pm0.024$ |
| *Herschel*-PACS-160 | 153.9 | $1.423\pm0.141$ |
| *Herschel*-SPIRE-250 | 242.8 | $1.430\pm0.020$ |
| *Herschel*-SPIRE-350 | 340.9 | $1.152\pm0.007$ |
| *Herschel*-SPIRE-500 | 482.3 | $0.735\pm0.002$ |
| JCMT-SCUBA-450 | 450 | $0.310\pm0.072$ |
| JCMT-SCUBA2-450 | 450 | $<1.082$ |
| JCMT-SCUBA-850 | 850 | $0.226\pm0.033$ |
| JCMT-SCUBA2-850 | 850 | $0.187\pm0.002$ |
| SMA-225 GHz | 1,332.41 | $(75.2\pm0.4)E-3^*$ |
| SMA-341 GHz | 878.2 | $(150.4\pm5.7)E-3^*$ |
| Toruń-OCRA-p | 1E4 | $(5.7\pm1.7)E-3$ |
| VLA-1.5 GHz | 2E5 | $< 1.52E-3$ |
| VLA-5 GHz | 6E4 | $1.34E-3$ |
| VLA-8 GHz | 3.7E4 | $1.53E-3$ |

**Figure ED1 | Spectral energy distribution of CK Vul.** Flux measurements and 3σ upper limits were collected from the literature and include data points obtained with the SMA in this study – as listed in Extended Data Table 4 and described in Supplementary Information. Best fit Planck function and modified grey-body spectral distributions are also shown.

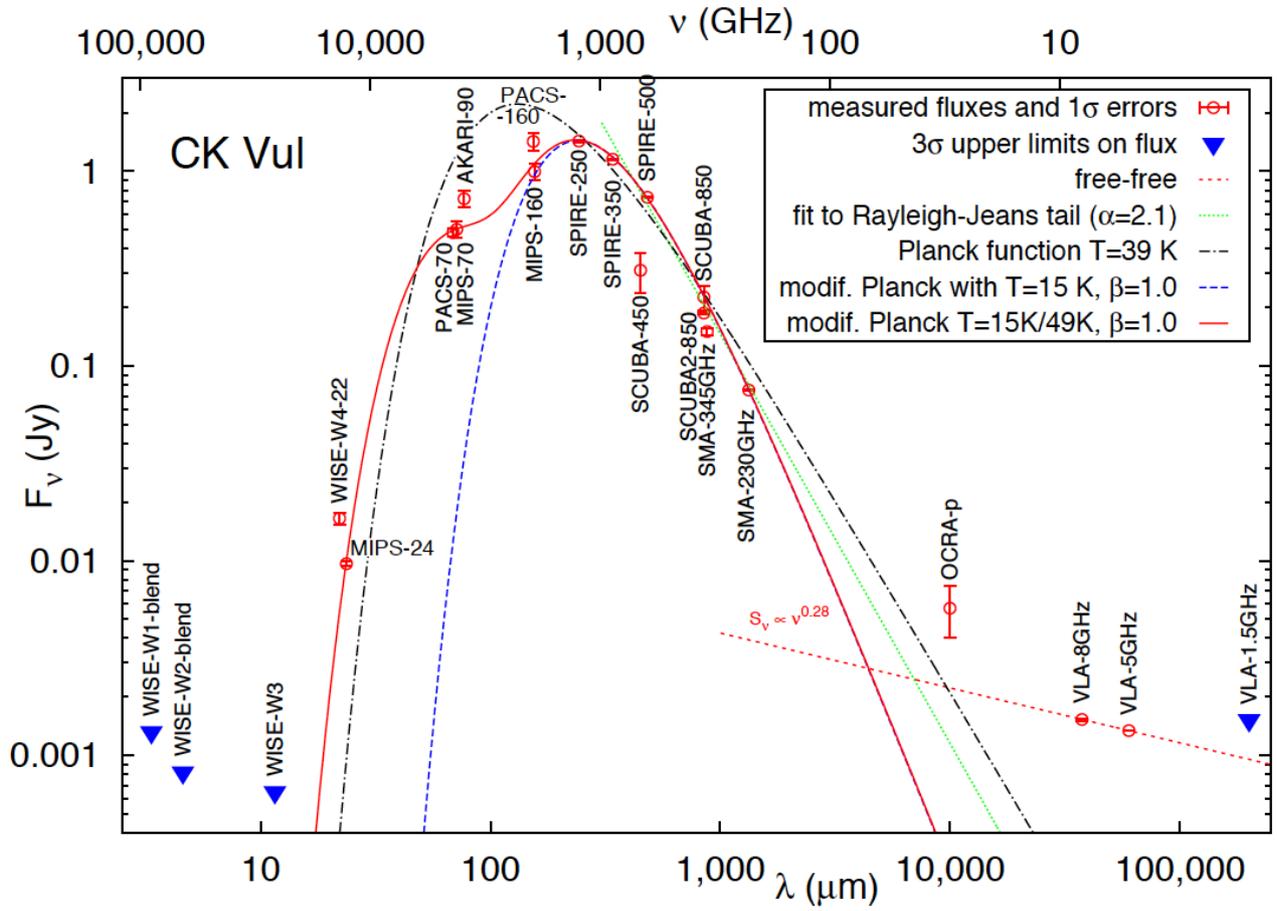